%% file: template.tex
\documentclass[preprint]{vgtc}                          

\graphicspath{{figures/}{pictures/}{images/}{./}} 

\usepackage{times}                     

\usepackage{tabu}                      
\usepackage{booktabs}                  
\usepackage{lipsum}                    
\usepackage{mwe}                       

\usepackage{mathptmx}                  

\usepackage[most]{tcolorbox}
\usepackage{xcolor}
\usepackage{xspace}
\usepackage{multicol}
\usepackage{subcaption}
\usepackage{multirow}
\usepackage{longtable}
\usepackage{balance}

\newcommand{\eg}{\textit{e.g.}}
\newcommand{\ie}{\textit{i.e.}}
\newcommand{\etal}{\textit{et~al.}}

\newcommand{\revise}[1]{\textcolor{black}{#1}}

\newcommand{\review}[1]{\textcolor{brown}{[#1]}\xspace}
\renewcommand{\review}[1]{}

\definecolor{mybrown}{HTML}{EEE0DA}
\definecolor{myred}{RGB}{255, 226, 221}
\definecolor{myorange}{RGB}{250, 222, 201}
\definecolor{mygreen}{RGB}{219, 237, 219}
\definecolor{mypink}{RGB}{245, 224, 233}
\definecolor{mypurple}{RGB}{232, 222, 238}
\definecolor{myblue}{RGB}{211, 229, 239}
\definecolor{myyellow}{RGB}{253, 236, 200}
\definecolor{mygrey}{RGB}{226, 226, 226}
\definecolor{mylightgrey}{RGB}{240, 240, 240}
\definecolor{navy}{RGB}{52, 85, 139}

\definecolor{BrickRed}{rgb}{0.8, 0.25, 0.33}
\definecolor{ForestGreen}{rgb}{0.13, 0.55, 0.13}
\definecolor{Brown}{rgb}{0.55, 0.27, 0.07}

\newcommand{\humanc}{\tcbox[on line, colback=myred, colframe=white, boxsep=0pt, left=2pt,right=2pt,top=2pt,bottom=3pt,arc=2pt,boxrule=0pt]{human-creator}\xspace}
\newcommand{\humancs}{\tcbox[on line, colback=myred, colframe=white, boxsep=0pt, left=2pt,right=2pt,top=2pt,bottom=3pt,arc=2pt,boxrule=0pt]{human-creators}\xspace}

\newcommand{\humano}{\tcbox[on line, colback=myorange, colframe=white, boxsep=0pt, left=2pt,right=2pt,top=2pt,bottom=1pt,arc=2pt,boxrule=0pt]{human-optimizer}\xspace}
\newcommand{\humanos}{\tcbox[on line, colback=myorange, colframe=white, boxsep=0pt, left=2pt,right=2pt,top=2pt,bottom=1pt,arc=2pt,boxrule=0pt]{human-optimizers}\xspace}

\newcommand{\humanonb}{{human-optimizer}\xspace}
\newcommand{\humanosnb}{{human-optimizers}\xspace}

\newcommand{\humana}{\tcbox[on line, colback=myyellow, colframe=white, boxsep=0pt, left=2pt,right=2pt,top=2pt,bottom=3pt,arc=2pt,boxrule=0pt]{human-assistant}\xspace}
\newcommand{\humanas}{\tcbox[on line, colback=myyellow, colframe=white, boxsep=0pt, left=2pt,right=2pt,top=2pt,bottom=3pt,arc=2pt,boxrule=0pt]{human-assistants}\xspace}

\newcommand{\humanr}{\tcbox[on line, colback=mypink, colframe=white, boxsep=0pt, left=2pt,right=2pt,top=2pt,bottom=3pt,arc=2pt,boxrule=0pt]{human-reviewer}\xspace}
\newcommand{\humanrs}{\tcbox[on line, colback=mypink, colframe=white, boxsep=0pt, left=2pt,right=2pt,top=2pt,bottom=3pt,arc=2pt,boxrule=0pt]{human-reviewers}\xspace}

\newcommand{\humanrnb}{{human-reviewer}\xspace}
\newcommand{\humanrsnb}{{human-reviewers}\xspace}

\newcommand{\aia}{\tcbox[on line, colback=mypurple, colframe=white, boxsep=0pt, left=2pt,right=2pt,top=2pt,bottom=3pt,arc=2pt,boxrule=0pt]{AI-assistant}\xspace}
\newcommand{\aias}{\tcbox[on line, colback=mypurple, colframe=white, boxsep=0pt, left=2pt,right=2pt,top=2pt,bottom=3pt,arc=2pt,boxrule=0pt]{AI-assistants}\xspace}

\newcommand{\aianb}{{AI-assistant}\xspace}
\newcommand{\aiasnb}{{AI-assistants}\xspace}

\newcommand{\aic}{\tcbox[on line, colback=myblue, colframe=white, boxsep=0pt, left=2pt,right=2pt,top=2pt,bottom=3pt,arc=2pt,boxrule=0pt]{AI-creator}\xspace}
\newcommand{\aics}{\tcbox[on line, colback=myblue, colframe=white, boxsep=0pt, left=2pt,right=2pt,top=2pt,bottom=3pt,arc=2pt,boxrule=0pt]{AI-creators}\xspace}

\newcommand{\aicnb}{{AI-creator}\xspace}
\newcommand{\aicsnb}{{AI-creators}\xspace}

\newcommand{\air}{\tcbox[on line, colback=mygreen, colframe=white, boxsep=0pt, left=2pt,right=2pt,top=2pt,bottom=3pt,arc=2pt,boxrule=0pt]{AI-reviewer}\xspace}

\newcommand{\aio}{\tcbox[on line, colback=mybrown, colframe=white, boxsep=0pt, left=2pt,right=2pt,top=2pt,bottom=1pt,arc=2pt,boxrule=0pt]{AI-optimizer}\xspace}

\newcommand{\na}{\tcbox[on line, colback=mylightgrey, colframe=white, boxsep=0pt, left=2pt,right=2pt,top=2pt,bottom=3pt,arc=2pt,boxrule=0pt]{N/A}\xspace}

\onlineid{0}

\vgtccategory{Research}

\vgtcinsertpkg

\title{Reflection on Data Storytelling Tools in the Generative AI Era from the Human-AI Collaboration Perspective}

\author{Haotian Li\thanks{e-mail: haotian.li@microsoft.com} \\
 \scriptsize Microsoft Research Asia 
\and Yun Wang\thanks{e-mail: wangyun@microsoft.com}\\ %
        \scriptsize Microsoft Research Asia %
\and Huamin Qu\thanks{e-mail: huamin@cse.ust.hk}\\ %
     \scriptsize The Hong Kong University of Science and Technology %
     }

\abstract{
    \input{sections/00-Abstract}
} 


\begin{document}

\maketitle

\input{sections/01-Introduction-v2}

\input{sections/02-Related}
\input{sections/03-Method}
\input{sections/04-01-Results}
\input{sections/04-02-Insights}

\input{sections/05-Discussion}
\acknowledgments{
The authors thank Leixian Shen and all reviewers for suggestions.}

\bibliographystyle{abbrv-doi}

\bibliography{main,chi24}

\input{sections/07-Appendix}

\end{document}

%% file: sections/00-Abstract.tex
Human-AI collaborative tools attract attentions from the data storytelling community to lower the expertise barrier and streamline the workflow.
The recent advance in large-scale generative AI techniques, \eg, large language models (LLMs) and text-to-image models, has the potential to enhance data storytelling with their power in visual and narration generation.
After two years since these techniques were publicly available, it is important to reflect our progress of applying them and have an outlook for future opportunities.
To achieve the goal, we compare the collaboration patterns of the latest tools with those of earlier ones using a dedicated framework for understanding human-AI collaboration in data storytelling.
Through comparison, we identify consistently widely studied patterns, \eg, human-creator + AI-assistant, and newly explored or emerging ones, \eg, AI-creator + human-reviewer.
The benefits of these AI techniques and implications to human-AI collaboration are also revealed.
We further propose future directions
to hopefully ignite innovations.




%% file: sections/01-Introduction-v2.tex
\section{Introduction}\label{sec:introduction}
Data storytelling is considered one of the major research directions in visualization research.
To facilitate telling appealing and effective stories, researchers have spent considerable efforts to build AI-powered tools with different strategies of human-AI collaboration~\cite{li2024we}.
In these tools, AI collaborators are often powered by heuristic-based methods, traditional machine learning models, or smaller-scale deep learning models~\cite{wu2021ai4vis}.


Compared to previous techniques used for AI collaborators, recently emerging large-scale generative AI models, including the text-to-image models and large language models (LLMs), can achieve better performance on various data storytelling-related tasks, such as data analysis and text generation, and enhance human-AI communication through conversational interactions.
These advantages highlight their potentials to be game-changers in the research area of human-AI collaboration for data storytelling, including improving the experience of working with AI and diversifying the collaboration patterns between humans and AI~\cite{li2024we}.
After two years of the public release of these large-scale
generative models, this is a critical moment to reflect on how this research discipline progresses, identify the benefits of those powerful techniques comprehensively, and propose future research opportunities.
\revise{To achieve the goal, we compared the human-AI collaboration patterns of 27 latest tools in the generative AI era with those of
earlier tools, based on the \textit{roles} of human and AI and the \textit{stages} where they collaborate in the data storytelling workflow~\cite{li2024we}.
Through the comparison, we find patterns that are still widely studied in the new era,} such as the mode where AI assists human creators, and newly explored or emerging patterns, such as humans acting as reviewers of AI-generated content.
We then discuss the implications for human-AI collaboration, such as the advantages of these large-scale generative AI models, the necessity and suitable scenarios for human-AI collaborative or fully automatic tools, and noteworthy research opportunities, such as the dependency of roles in different stages.
Through discussions on the progress and future opportunities, we hope this paper can serve as a cornerstone for investigating the collaboration between humans and large-scale generative AI.


%% file: sections/02-Related.tex
\section{Related Surveys}\label{sec:related_work}

To review the progress of data storytelling tools, 
the first group of surveys focuses on how tools support narrative techniques, such as data story structures~\cite{tong2018storytelling} and components~\cite{shen2025reflecting}.
The second group focuses on the application of automation techniques, such as AI, in data storytelling tools.
Chen~\etal~\cite{chen2023does} categorized AI-powered tools into manual, mixed-initiative, and automatic types.
Li~\etal~\cite{li2024we} proposed a fine-grained framework to explicitly characterize the collaboration between human and AI through collaboration stages and roles.
Recently, 
He~\etal~\cite{he2024leveraging} reviewed generative AI model usage in tools based on tasks, such as insight identification.
\revise{However, its taxonomy does not consider how humans work with AI models.
Under the second category, our paper aims to identify insights about how the research in human-AI collaboration for data storytelling is advanced in the era of generative AI models.
As a sequel to the work by Li~\etal~\cite{li2024we}, it refreshes our understanding through comparing the human-AI collaboration patterns of the latest tools with those of earlier tools and summarizes future opportunities.}




%% file: sections/03-Method.tex
\section{Method}\label{sec:method}
This section introduces our research method of collecting tools and analyzing the collaboration patterns.
For brevity, we denote the tools after Jun. 2023 as the \textit{latest tools} in the generative AI era and the tools before as \textit{earlier tools}. 

\subsection{Tool Collection}
\revise{To collect the corpus, we searched for papers published between Jun. 2023 and Dec. 2024 in related premier HCI venues, including CHI, UIST, and IUI, and visualization venues, including  VIS, PacificVis, EuroVis, and TVCG, with keywords ``narrative visualization'' and ``data storytelling''.
We also supplemented the corpus with papers that were reviewed in a latest related survey~\cite{shen2025reflecting}. 
Following Li~\etal~\cite{li2024we}, we then filtered out the papers that are not for storytelling purpose with infomation visualization or do not describe tools.
Finally, we collected 27 tools, as shown in Table~\ref{tab:tool_list}.
As a preliminary investigation, this paper focuses on the tools in the academia since these tools are often described with more details, such as design rationales and usage scenarios, making it easier to identify the high-level collaboration patterns and underlying reasons.
It will be valuable to expand the scope to commercial tools for more comprehensive understanding.
}

After a careful consideration, we compared human-AI collaboration patterns of the latest tools after Jun. 2023 with those of earlier tools to study the research trend of human-AI collaborative tool in the era of generative AI.
First, 
the public release dates of two representative large-scale generative AI techniques, ChatGPT and Stable Diffusion, are Nov. 30, 2022\footnote{\href{https://openai.com/index/chatgpt/}{https://openai.com/index/chatgpt/}} and Aug. 22, 2022\footnote{\href{https://stability.ai/news/celebrating-one-year-of-stable-diffusion}{https://stability.ai/news/celebrating-one-year-of-stable-diffusion}}, respectively.
Considering the delay led by tool development and paper publishing, the research tools to apply them and other similar techniques might only be available in mid-2023. 
Then, we examined the tools collected by Li~\etal~\cite{li2024we}.
These tools were published by Jun. 2023 and did not have extensive application of these models.
Our collected tools further verify the decision: there were pioneering ones that apply text-to-image models~\cite{xiao2023let} and LLMs~\cite{sultanum2023datatales, shen2023dataplayer, lin2023inksight} published in VIS 2023 in Oct. 2023.

\input{tables/paper_list}

\begin{figure*}
        \centering
        \begin{subfigure}[b]{0.24\linewidth}
            \centering
            \includegraphics[width=\textwidth]{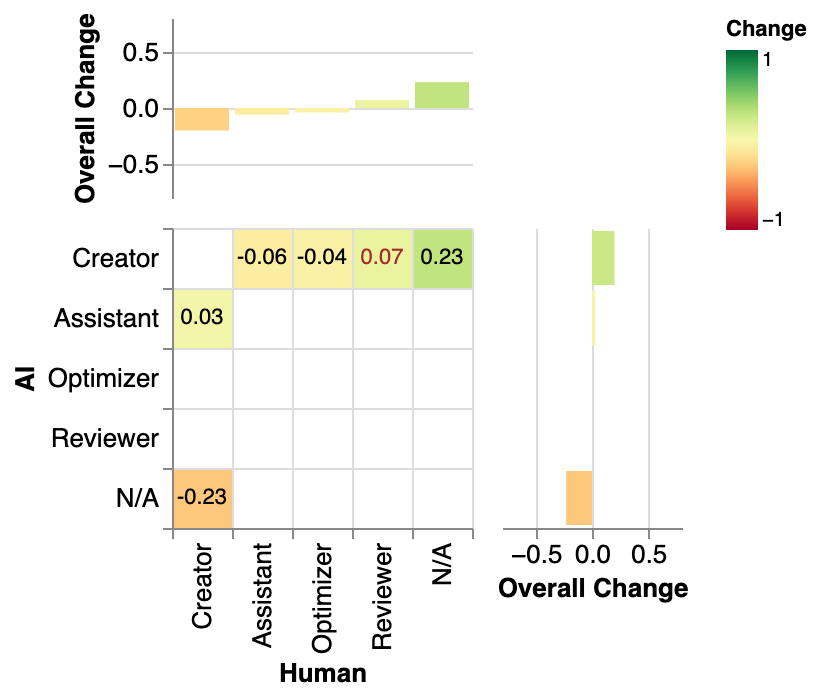}
            \caption{Analysis}  
        \end{subfigure}
        \begin{subfigure}[b]{0.24\linewidth}  
            \centering 
            \includegraphics[width=\textwidth]{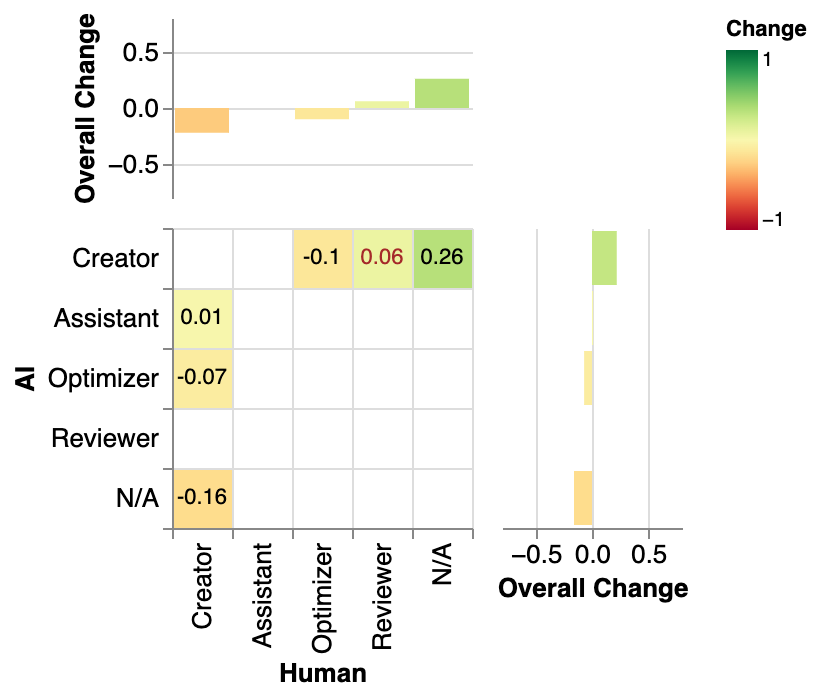}
            \caption{Planning}    
        \end{subfigure}
        \begin{subfigure}[b]{0.24\linewidth}   
            \centering 
            \includegraphics[width=\textwidth]{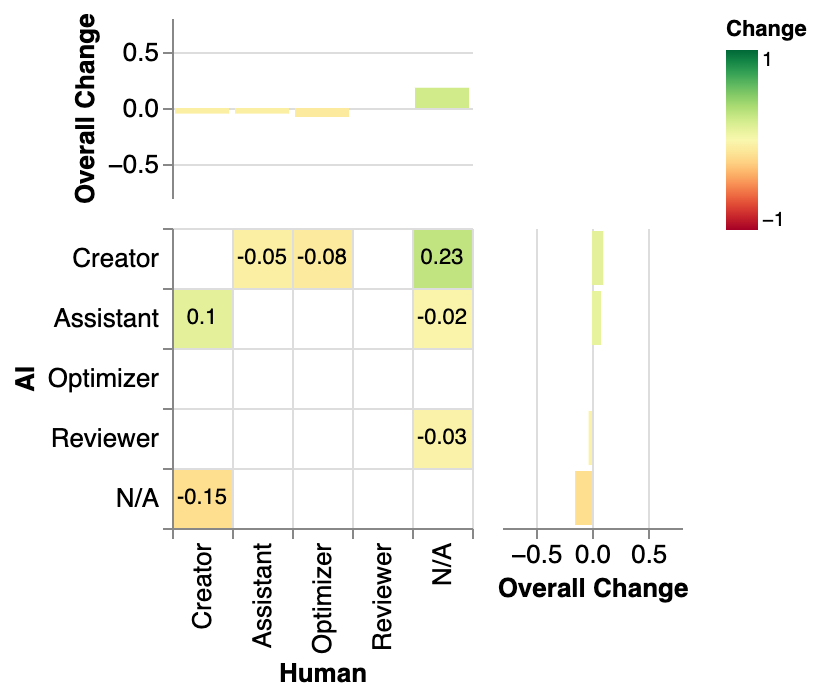}
            \caption{Implementation}    
        \end{subfigure}
        \begin{subfigure}[b]{0.24\linewidth}   
            \centering 
            \includegraphics[width=\textwidth]{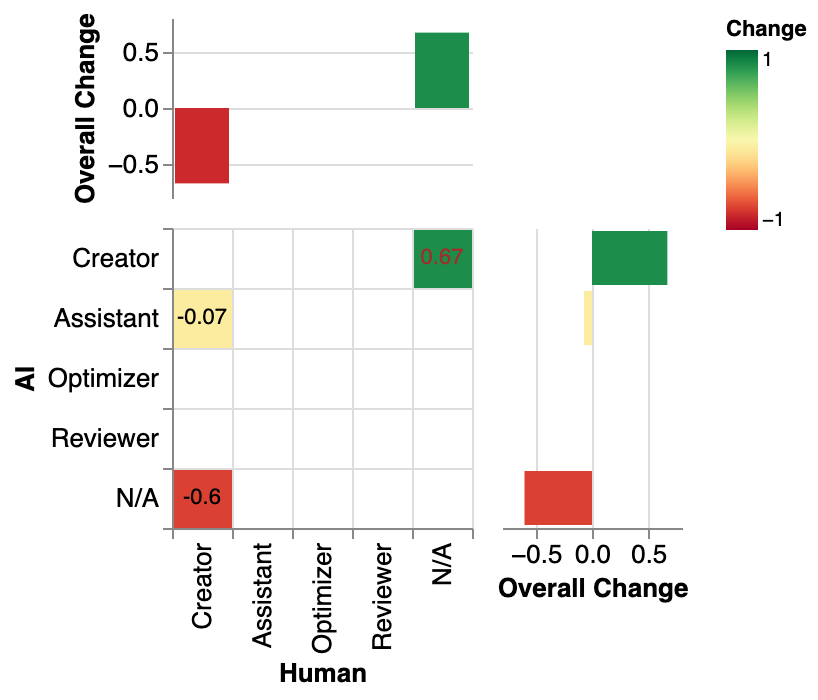}
            \caption{Communication}  
        \end{subfigure}
        \vspace{-1em}
        \caption
        {This figure shows the change of human-AI collaboration pattern frequencies between the latest and earlier tools in four stages. 
        In a subfigure, the color of blocks and numbers in the heatmap encodes the change of pattern frequencies in the stage. A pattern is represented by a human role on the horizontal axis and an AI role on the vertical axis. \textcolor{ForestGreen}{Green} and positive numbers mean that the pattern is more frequent in the latest tools while \textcolor{BrickRed}{red} and negative numbers indicate that the pattern is less frequent. A number in \textcolor{Brown}{brown} means that the pattern is newly identified in the latest tools. An empty white cell means that the pattern has not emerged in any tools.}
        \label{fig:stage_change}
        \vspace{-1em}
    \end{figure*}

\subsection{Pattern Coding}
After collecting all tools, we coded their human-AI collaboration patterns with the framework proposed by Li~\etal~\cite{li2024we}.
The framework has two dimensions: \textbf{the roles of human and AI collaborators} 
(\ie, \humanc and \aic to complete the majority of tasks in each stage, \humana and \aia to help creators on designated tasks in a stage, \humano and \aio to improve the quality of the output of one task or the entire stage autonomously, and \humanr and \air to provide suggestions for the output from a task or stage) 
and \textbf{their collaboration stages} (\ie, \textit{analysis} stage to derive data insights and knowledge from datasets, \textit{planning} stage to decide how to convey analytical findings through a data story, \textit{implementation} stage to create data story content following the plan, and \textit{communication} stage to share data stories directly with audiences).
One author coded all tools after understanding the coding principles and the coded tools by Li~\etal~comprehensively.
During the coding process, the results were also calibrated with previously coded tools as references.
The final results are presented in Table~\ref{tab:tool_list} and an online browser: \url{https://human-ai.notion.site/data-storytelling-dec-2024}.
\revise{In Table 1, to help readers compare the tools in two papers more conveniently, the order of tools aligns with the sorting method used in Li~\etal~\cite{li2024we}, which is based on the stages each tool covers.
In each stage, the leading role (\ie, creator) in one stage will be shown first.}
To facilitate comparison, we computed and visualized the changes in the \textit{relative frequencies}\footnote{All frequencies refer to relative frequencies to accommodate the different numbers of tools in each stage in earlier and the latest tools.} of collaboration patterns in Fig.~\ref{fig:stage_change}.
The frequency of a pattern $p$ in a stage $s$ is computed as $f = Occurrence_{p \textrm{ in } s}/NumTool_{s}$ and the frequency change between the latest tools and earlier tools is $f_{latest}-f_{earlier}$.
In Fig.~\ref{fig:stage_change}, a positive change  means that \revise{the pattern is gaining more attention} while a negative one shows the frequency is declining.
\revise{Our supplementary material also present the relative frequency of collaboration patterns among earlier and the latest tools for our readers to understand what patterns are widely studied.}

We applied the framework for three-fold reasons.
First, the framework distinguishes the collaboration patterns of human and AI at a fine granularity.
Furthermore, the framework was summarized based on an interview study~\cite{li2023ai} that envisions potential AI application.
Evaluating the alignment between existing tools and user expectations can help us reflect on current practices and identify future opportunities.
Another advantage is that the framework is technique-agnostic, allowing us to compare collaboration patterns without interference led by differences in techniques.
Combining the results of ours and Li~\etal~\cite{li2024we}, we notice \textit{a considerable growth in the number of data storytelling tools in 2024}.
Compared to the 11 tools published in 2023, the 20 tools in 2024 nearly double the previous number, reaching a historic high.
Furthermore, 16 out of the 27 latest tools apply generative AI models (Table~\ref{tab:tool_list}).
These models are most frequently used in the implementation stage (15/27), following up by the analysis (5/27), the planning (4/27), and the communication stages (3/27).
The findings reveal that \textit{large-scale generative AI models has been widely applied across the storytelling workflow}, supporting our motivation.


%% file: tables/paper_list.tex
\begingroup
\renewcommand{\arraystretch}{0.8}

\newcolumntype{H}{>{\setbox0=\hbox\bgroup}c<{\egroup}@{}}

\renewcommand{\humanc}{\tcbox[on line, colback=myred, colframe=white, boxsep=0pt, left=2pt,right=2pt,top=1pt,bottom=1pt,arc=2pt,boxrule=0pt]{H-C}\xspace}
\renewcommand{\humano}{\tcbox[on line, colback=myorange, colframe=white, boxsep=0pt, left=2pt,right=2pt,top=1pt,bottom=1pt,arc=2pt,boxrule=0pt]{H-O}\xspace}
\renewcommand{\humana}{\tcbox[on line, colback=myyellow, colframe=white, boxsep=0pt, left=2pt,right=2pt,top=1pt,bottom=1pt,arc=2pt,boxrule=0pt]{H-A}\xspace}
\renewcommand{\humanr}{\tcbox[on line, colback=mypink, colframe=white, boxsep=0pt, left=2pt,right=2pt,top=1pt,bottom=1pt,arc=2pt,boxrule=0pt]{H-R}\xspace}

\renewcommand{\aia}{\tcbox[on line, colback=mypurple, colframe=white, boxsep=0pt, left=2pt,right=2pt,top=1pt,bottom=1pt,arc=2pt,boxrule=0pt]{A-A}\xspace}
\renewcommand{\aic}{\tcbox[on line, colback=myblue, colframe=white, boxsep=0pt, left=2pt,right=2pt,top=1pt,bottom=1pt,arc=2pt,boxrule=0pt]{A-C}\xspace}
\renewcommand{\air}{\tcbox[on line, colback=mygreen, colframe=white, boxsep=0pt, left=2pt,right=2pt,top=1pt,bottom=1pt,arc=2pt,boxrule=0pt]{A-R}\xspace}
\renewcommand{\aio}{\tcbox[on line, colback=mybrown, colframe=white, boxsep=0pt, left=2pt,right=2pt,top=1pt,bottom=1pt,arc=2pt,boxrule=0pt]{A-O}\xspace}

\renewcommand{\na}{\tcbox[on line, colback=mylightgrey, colframe=white, boxsep=0pt, left=2pt,right=2pt,top=1pt,bottom=1pt,arc=2pt,boxrule=0pt]{N/A}\xspace}

\begin{table}[]
    \caption{This table presents the human-AI collaboration patterns
    and the applied generative AI techniques in our collected tools. The collaborators' roles are shown in abbreviation: \humanc, \humana, \humano, \humanr are the creator, assistant, optimizer, and reviewer roles performed by humans; \aic, \aia, \aio, \air are the corresponding roles by AI. 
    The techniques are denoted with the icons: * refers to LLMs
    and \# refers to text-to-image models.}
    \small
    \centering
    \begin{tabular}{Hp{0.5cm}p{1.5cm}p{1.5cm}p{1.5cm}p{1.5cm}}
    \toprule
         \textbf{Year} & \textbf{Tool} & \textbf{Analysis} & \textbf{Planning} & \textbf{Implementation} & \textbf{Communication} \\\midrule
        2023 & \cite{yao2023designing} & \na & \na & \humanc & \na \\
        2024 & \cite{zhou2024epigraphics} & \na & \na & \humanc \aia$^{*\#}$ & \na \\ 
        2024 & \cite{chen2024beyond} & \na & \na & \humanc \aia$^{*\#}$ & \na \\
        2024 & \cite{brehmer2024data} & \na & \na & \humanc \aia & \na \\ 
        2024 & \cite{shen2024authoring} & \na & \na & \humanc \aia & \na \\ 
        2024 & \cite{cai2024linking} & \na & \na & \humanc \aia & \na \\ 
        2023 & \cite{sultanum2023datatales} & \na & \na & \aic$^{*}$ \humano & \na \\ 
        2024 & \cite{tong2024vistellar} & \na & \humanc & \humanc \aia & \na \\
        2024 & \cite{wang2024wonderflow} & \na & \humanc & \humanc \aia & \na \\
        2024 & \cite{femi2024visconductor} & \na & \humanc & \humanc & \humanc \aia \\
        2023 & \cite{shen2023dataplayer} & \na & \humanc & \aic$^{*}$ & \na \\
        2024 & \cite{shen2024dataplaywright} & \na & \humanc & \aic$^{*}$ \humano & \na \\
        2023 & \cite{xiao2023let} & \humanc \aia & \na & \humanc \aia$^{\#}$ & \na \\
        2023 & \cite{lin2023inksight} & \humanc & \na & \aic$^{*}$ \humano & \na \\
        2024 & \cite{wang2024outlinespark} & \humanc & \humanc \aia$^{*}$ & \aic$^{*}$ \humano & \na \\
        2024 & \cite{zhao2024leva} & \humanc \aia$^{*}$ & \humanc & \aic$^{*}$ & \na \\
        2024 & \cite{li2024coinsight} & \humanc \aia & \humanc \aia & \aic & \na \\
        2024 & \cite{shen2024datadirector} & \aic$^{*}$ & \aic$^{*}$ & \aic$^{*}$ & \na \\
        2024 & \cite{shi2024understanding} & \aic & \aic & \aic & \na \\
        2024 & \cite{ying2024reviving} & \aic$^{*}$ & \aic$^{*}$ & \aic$^{*}$ & \aic$^{*}$ \\
        2024 & \cite{lee2024sportify} & \aic$^{*}$ & \aic$^{*}$ & \aic$^{*}$ & \aic$^{*}$ \\
        2024 & \cite{andrews2024aicommentator} & \aic & \aic & \aic$^{*}$ & \aic \\
        2023 & \cite{chen2023calliope} & \aic \humano & \aic \humano & \aic \humano & \na \\
        2024 & \cite{kim2024dg} & \aic \humano & \aic \humano & \aic$^{*}$ \humano & \na \\
        2024 & \cite{cheng2024snil} & \aic$^{*}$ \humano & \aic \humano & \aic$^{*}$ \humano & \na \\
        2023 & \cite{wu2023socrates} & \aic \humanr & \aic \humanr & \aic & \na \\
        
        2024 & \cite{han2024deixis} & \na & \na & \na & \humanc \aia$^{*}$ \\
        \bottomrule
    \end{tabular}
    \vspace{-1em}
    \label{tab:tool_list}
\end{table}

\endgroup


%% file: sections/04-01-Results.tex
\section{Findings}\label{sec:finding}
\revise{This section presents insights into the evolving collaboration patterns studied within the community, including consistently widely studied patterns, newly explored patterns, and those gaining increased attention.
It is important to note that changes in patterns should not be interpreted as indicators of success or failure.
They only reflect shifts in research interest.}

%% file: sections/04-02-Insights.tex
\subsection{Which Patterns Are Still Widely Studied?}\label{sec:finding_remain}
The most frequent patterns involving both humans and AI in earlier tools are \humanc + \aia\footnote{The plus sign indicates the two roles collaborate in a specific stage.} and \aic + \humano since they allow humans to control the content while lowering the barrier of creation through automation~\cite{li2024we}.
In the latest tools, \textit{their frequency only has limited 
fluctuation, suggesting that they still attract considerable research interests \revise{(see Figs. 1 (a)-(c))}.}
\revise{As the characteristics of the two patterns have been extensively discussed~\cite{li2024we}, we focus on the advancements brought by generative AI models when applying these patterns.}

First, large-scale generative AI models can improve the output and experience of human-AI collaboration.
They can \textit{increase AI collaborators' output space significantly}.
Several latest tools~\cite{zhou2024epigraphics, chen2024beyond, xiao2023let} 
apply \aias to create visual components and provide design inspirations to support \humancs for authoring static or animated infographics.
The similar design exists in earlier tools, such as
DataQuilt~\cite{zhang2020dataquilt}.
However, earlier \aiasnb can only support re-using existing visual elements, greatly limiting the potential space of creation.
The AI collaborators powered by retrieval-based algorithms also suffer from the issue, 
such as the \aic in Retrieve-then-Adapt~\cite{qian2020retrieve}. 
Compared to earlier work, the latest tools powered by generative AI models are more flexible to create new visual elements beyond existing examples.
Additionally, \textit{the interaction between collaborators can be more natural and closer to humans' habits}.
In addition to prompting~\cite{xiao2023let, zhao2024leva}, OutlineSpark~\cite{wang2024outlinespark} allows users to input a structured outline which guides \aicnb to generate slides.
In Data Playwright~\cite{shen2024dataplaywright}, users write annotated narrations to specify expected animation effect and timing along with narrations.
In both tools, users' intent are conveniently expressed to enhance their agency, and their efforts are reduced through more AI automation.
The last advantage is \textit{output quality improvement}, such as enhancing text generation~\cite{lin2023inksight, kim2024dg}.


\revise{Second, \textit{large-scale generative AI models have the potential to enable innovative application of AI collaborators.}}
The previous usage of \aia in the communication stage was to support individual users in live presentations, such as retrieving and presenting charts based on sketch~\cite{lee2013sketchstory}.
Building on LLMs' ability to understand and generate texts, Han and Isaacs~\cite{han2024deixis} proposed an innovative usage case of \aianb to 
record the communication process between \humancs around charts and create an interactive data document based on the recording.
The work demonstrates a new collaborative data story creation mode, where \textit{humans freely express their insights and AI documents and organizes humans' ideas into a more structured data story}.
It is interesting to examine the benefit of this collaboration mode and explore other potentials.
For example, what about asking AI to interview humans and write data stories or generating data videos like talk shows based on humans' conversations around data?
Furthermore, \textit{from the human-AI collaboration perspective, this \aianb has a notable difference with others}.
It serves as a third-party collaborator to document multiple humans' communication, while not collaborating with individual humans and contributing to their creation or presentation.
We should further investigate the potentials of human-AI collaboration to support multiple humans' work in data storytelling~\cite{chevalier2018analysis}.
\revise{The provocation also aligns with a finding from Fig.~\ref{fig:stage_change}.
In Fig.~\ref{fig:stage_change}, many underexplored patterns are without any human- or AI-creators, such as \aio + \humano.
With future tools following these patterns, AI and humans will often collaborate to support third-party creators who do not use the tool.}


\subsection{What New Patterns Have Been Explored?}\label{sec:finding_new}
\revise{From Figs. 1 (a), (b), and (d)}, we identified new collaboration patterns that were explored: \textit{\humanrs in the analysis and the planning stages and \aics in the communication stage}.

The collaboration pattern of \aic + \humanr has been discussed as a potential strategy to enhance AI automation and human agency~\cite{li2024we}.
Compared to \humanos, \humanrsnb can control the data story with feedback on AI-created content and thus save the effort to revise the content manually.
Socrates~\cite{wu2023socrates} allows \humanrnb to provide feedback to the direction of data analysis (\eg, what attributes in datasets to be included in the data story) and the narrative structure (\eg, highlight a data fact using the strategy of contrasting it with another one).
Specifically, the \aicnb, powered by a heuristic-based algorithm, will generate questions and ask for \humanrnb to rate, compare, or select options. 
Then, the \aicnb can leverage the results to create a data story that aligns with the intent of \humanrnb.
Though they do not explicitly compare the efforts spent with their tool and another with \humanonb, \ie, Calliope~\cite{shi2020calliope}, their results show that the created stories with Socrates match the intent better, which also implies that humans might save efforts for further adjustment.
\textit{The study results preliminarily verify the advantages of \humanrnb to achieve more AI automation and more human agency than \humanonb, when collaborating with an \aicnb.}
\revise{In addition to the authors' suggestion of exploring free-form feedback, future research can further study \humanrsnb, such as understanding their pros and cons.}


Another new pattern is the application of \aics for communicating data stories automatically.
According to an interview study, data workers often prefer AI not to perform the creator role in the communication since they may diminish the underlying value of human-human communication, such as building trust~\cite{li2023ai}.
Therefore, we are particularly interested in the design of these \aicsnb in the latest tools.
Sportify~\cite{lee2024sportify} and AiCommentator~\cite{andrews2024aicommentator} leverage \aicsnb to communicate AI-identified data insights in sports games to humans.
Ying~\etal~\cite{ying2024reviving} and Shi~\etal~\cite{shi2024understanding} support chart and dataset understanding with \aicsnb who communicate insights to humans through AI-created animations and narrations.
Comparing these cases with the interview study, we notice that \textit{these \aicsnb communicates AI-created stories rather than presenting human-created stories on behalf of humans}.
It implies whether \aicsnb can be accepted by humans in the communication stage may depend on AI's roles in the previous stages.
If the story is fully created by the AI collaborator in previous stages and reflect little human opinions, the role of \aicsnb is more acceptable.
On the other hand, when the story carries intensive information from human storytellers, the usage of \aicsnb as a human proxy might raise concerns.
\revise{Furthermore, the observation also
indicates that \textit{the application of different roles across stages can be dependent on each other}, which needs further investigation.}







\subsection{Which Patterns Are Gaining More Attention?}\label{sec:finding_emerging}

\revise{In Fig. 1,} we notice that \textit{the \aic-only pattern attracts growing research interests},
where fully automatic tools across multiple stages emerge~\cite{ying2024reviving, lee2024sportify, shen2024datadirector, andrews2024aicommentator}.
We believe it is led by the increasing capability of large-scale generative AI models, such as the power of analyzing data~\cite{inala2024data} and generating text explanation~\cite{dibia2023lida}.
One might question if the trend contradicts with the argument that human-AI collaboration in data storytelling would gain more interests~\cite{li2024we}.
Furthermore, does the advancement of AI capability mean that we should try to automate everything with AI in data storytelling and thus the research on human-AI collaboration is no longer needed?

Inspired by the discussion where collaboration patterns can be affected by tool usage scenarios~\cite{li2024we},
we dived into the scenarios of these fully automatic tools to figure out why they are designed to be \aics-only.
In the usage scenarios of Sportify~\cite{lee2024sportify} and AiCommentator~\cite{andrews2024aicommentator}, sports fans issue targeted queries about matches for desired information, such as tactics or player performances.
Then, AI collaborators create tailored and casual data stories based on the queries and communicate them to humans automatically.
Compared to other scenarios, such as presenting data stories to the public, this casual and low-stake one 
\textit{emphasizes the need of story customization and real-time creation while tolerating that the quality might not be perfect}.


To meet the demands of customization and high responsiveness, \textit{the creation and communication of data stories can hardly be intervened by \humancs or \humanas}, who are likely to take longer time and higher financial cost to finish tasks~\cite{li2023ai}.
Traditional AI algorithms may also have limited flexibility for those highly customized data stories.
The emergence of large-scale AI models helps fill the gap due to their capability to process free-form input that describes audience goals and to generate and communicate data stories accordingly (Sec.~\ref{sec:finding_remain}).
They can act as \aics to automate the workflow.
Since the scenarios are casual and low-stake, careful examination and quality improvement by \humanos or \humanrs might not be needed.

The scenario of Live Charts~\cite{ying2024reviving} and the work by Shi~\etal~\cite{shi2024understanding} impose similar requirements to storytellers.
To facilitate chart understanding, the authors of these two papers propose to transform static charts into data stories with animations and narrations.
The generated data stories are intended to help users interpret their charts more quickly, while polished design is considered a lower priority.
Given the high cost of humans further, a fully automatic pipeline is more suitable.
Another similar audience-oriented scenario is to enhance text-only data story understanding by generating visuals~\cite{li2024we}, where automatic tools~\cite{hullman2013contextifier, shao2024narrative} are designed.

After examining these cases, we can attempt to answer the question about the necessity and potential future research direction of human-AI collaboration.
First, led by the outstanding capability of large-scale generative AI models, 
\textit{it is a natural and irresistible trend that the application of \aicsnb will be wider}.
To cope with the trend, we should actively experiment with these models~\cite{li2024we}
and learn how to apply them smartly, such as effective prompt design~\cite{shen2024datadirector}.
Furthermore, we should realize that \textit{the application of \aicsnb is not only related to their performance but the scenario as well}.
The scenarios of fully automatic tools share common features, including low-stake scenarios that require real-time generation of personalized stories.
Additionally, these scenarios typically do not involve human-human communication (Sec.~\ref{sec:finding_new}).
In other cases, fully automatic tools might not be suitable.
For example, in contrast to the aforementioned tools~\cite{lee2024sportify, andrews2024aicommentator},
SNIL~\cite{cheng2024snil} is designed for the scenario where data journalists create serious content for the public.
As a result, though SNIL can generate data articles for basketball games, it emphasizes the collaboration between \aicsnb and data journalists as \humanosnb.
Furthermore, in scenarios requiring creativity, the research of human-AI collaborative tools is also booming (\eg,~\cite{chen2024beyond, xiao2023let, zhou2024epigraphics}).
More research is needed to understand the relationship between the needs for different AI collaborators and the characteristics of scenarios.
Then, in those scenarios requiring human-AI collaboration, designing suitable collaboration patterns should be further studied.
Finally, \textit{though these tools appears lack of human-AI collaboration from the data storytelling perspective, they support human-AI collaboration from a broader angle}.
For example, AI-created data stories help sports fans to understand tactics~\cite{lee2024sportify} and assist novices to comprehend charts~\cite{ying2024reviving}.
To conclude our response to the questions, we believe that human-AI collaboration remains an important topic to investigate and there will be more opportunities to apply data storytelling for broader human-AI collaboration in data-related tasks.




%% file: sections/05-Discussion.tex

\section{Future Opportunities}


\textbf{Large-scale Generative AI Models as Universal Engines with Unified Interfaces.}
In addition to the benefits in Sec.~\ref{sec:finding_remain}, a more valuable characteristic of these generative AI models, especially LLMs, is to \textit{provide universal engines and interfaces for various roles and stages}.
Earlier tools often equip an AI collaborator with a specialized algorithm in one stage, \eg, fact mining and organization~\cite{li2023notable}.
Next, an interface, or ``shared representation''~\cite{heer2019agency} is required to support the communication among AI collaborators in different stages and even between AI and humans~\cite{li2024we}, \eg, chart or insight specification.
Existing research shows that LLMs can support these stages as different roles with appropriate prompt design~\cite{lee2024sportify}.
Furthermore, they can leverage natural languages as a unified interface for communication, with supplementary ways of programming languages.
Therefore, these models can reduce tool development cost led by specialized algorithms and release the constraint of communication channels. 
Researchers can apply them to investigate 
expected but overlooked patterns, \eg, \aio+\humano~\cite{li2024we} and \humanc + \air~\cite{li2023ai}, conveniently.
Furthermore, it is exciting to explore fluid human-AI collaboration with user-configured patterns.

\noindent \textbf{Guidelines for Applying AI Models.}
Though powerful, these large-scale generative AI models have downsides, such as long response time
and risks like hallucination.
As a result, their usage requires \textit{more research at a meta level to summarize when, where, and how they should be leveraged to achieve more benefit while mitigating their issues}.
For example, it is necessary to consider how to decompose tasks and design prompts~\cite{shen2024datadirector} and combine their usage with traditional AI algorithms to enhance the processing speed, accuracy, and reliability~\cite{li2025composing}.
A design trade-off discussed by Socrates authors~\cite{wu2023socrates} is whether to apply LLMs for extracting intent from free-form feedback input by \humanrsnb or only to support pre-defined input types for more direct preferences.
It is interesting to propose AI usage guidelines based on techniques, usage scenarios, and collaboration patterns.

\noindent \textbf{Necessity for Continuous Reflection.}
In the generative AI era, we are witnessing a fast growth of data storytelling tools.
The research progress should be reflected continuously and regularly to summarize lessons and identify opportunities.
For example, Li~\etal~only discussed the existence of the relationship between collaboration patterns and usage scenarios with an example of \aicsnb to support data story understanding~\cite{li2024we}.
Powered by generative AI, more \aicnb-only scenarios emerges, allowing us to 
summarize their common characteristics, such as the casual and audience-oriented nature (Sec.~\ref{sec:finding_emerging}).
These findings greatly enrich our knowledge about the effect of usage scenarios on tool design.
\revise{Furthermore, the advancement of large-scale generative AI models greatly diversifies methods to leverage them.
Users can finish various tasks in data storytelling with simple chat-based interfaces instead of carefully designed tools.
For example, with commercial chatbots powered by 
multi-modal LLMs, such as GPT-4o and Gemini, users can analyze data, create texts, generate visuals, and iterate designs by prompting models.
These observations advise that human-AI collaboration evolves with technical advancements.
As a result, we must refresh our understanding continuously to gather useful knowledge related to both academic and commercial tools and keep it up-to-date.}

%% file: sections/07-Appendix.tex


\begin{figure*}[h!]
    \centering
    \includegraphics[width=0.5\linewidth]{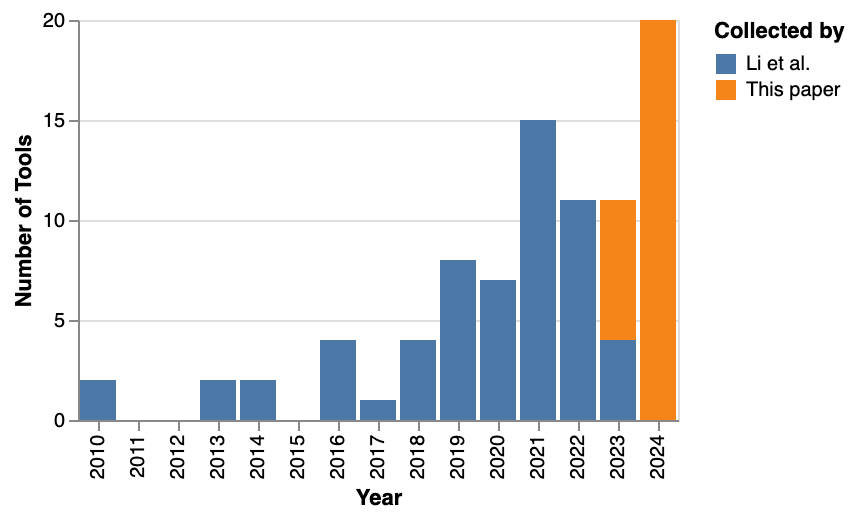}
    \caption{The number of data storytelling tools between 2010 and 2024}
    \label{fig:trend}
\end{figure*}

\begin{figure*}[h!]
        \centering
        \begin{subfigure}[b]{0.24\linewidth}
            \centering
            \includegraphics[width=\textwidth]{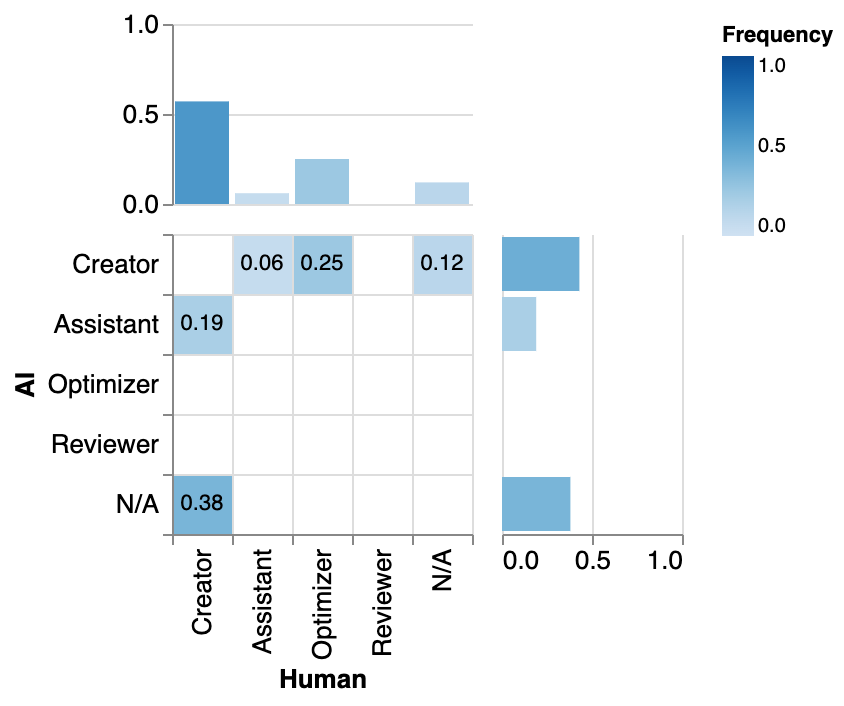}
            \caption{Analysis}  
        \end{subfigure}
        \begin{subfigure}[b]{0.24\linewidth}  
            \centering 
            \includegraphics[width=\textwidth]{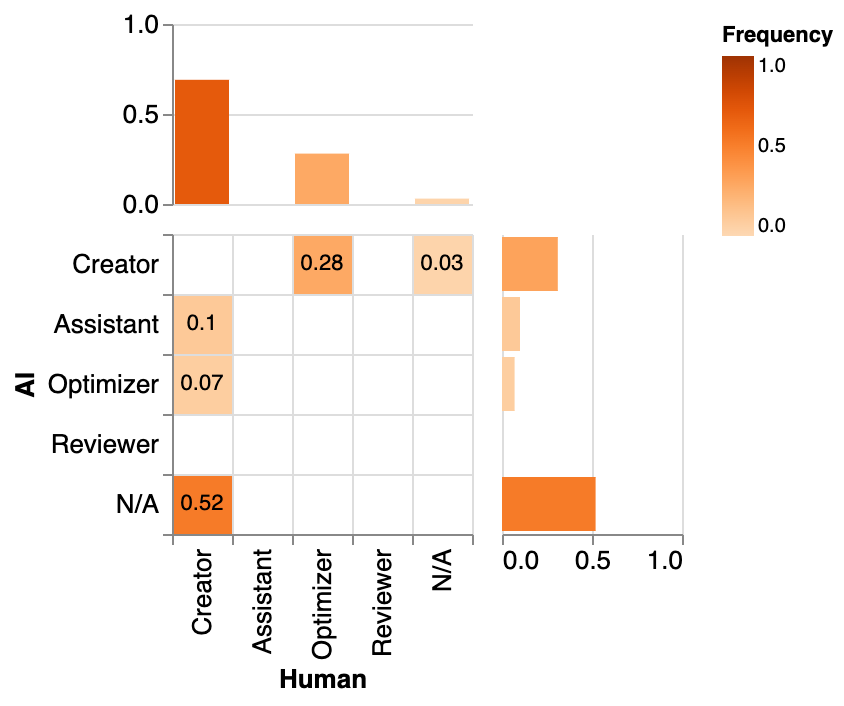}
            \caption{Planning}    
        \end{subfigure}
        \begin{subfigure}[b]{0.24\linewidth}   
            \centering 
            \includegraphics[width=\textwidth]{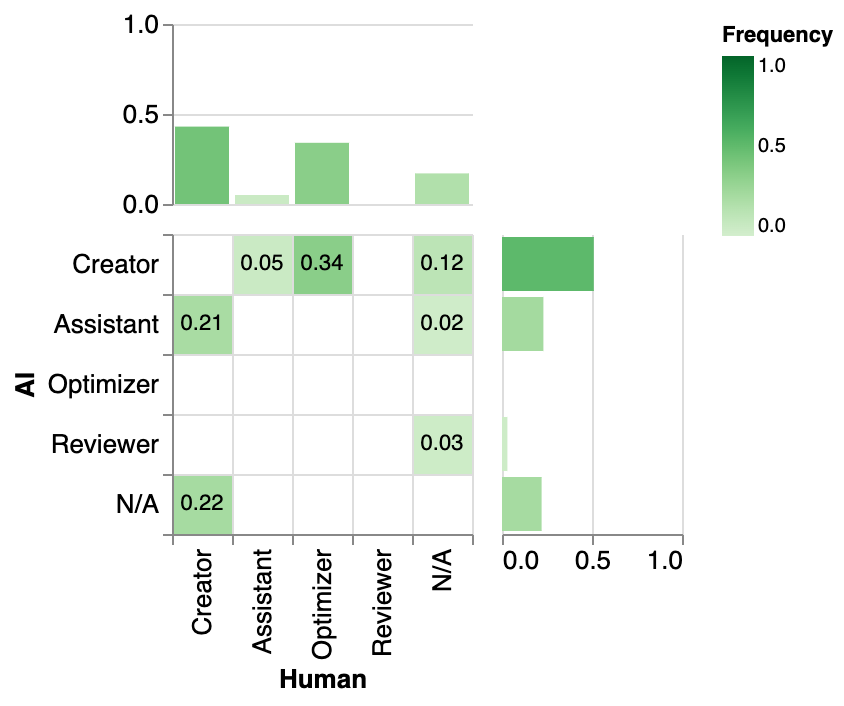}
            \caption{Implementation}    
        \end{subfigure}
        \begin{subfigure}[b]{0.24\linewidth}   
            \centering 
            \includegraphics[width=\textwidth]{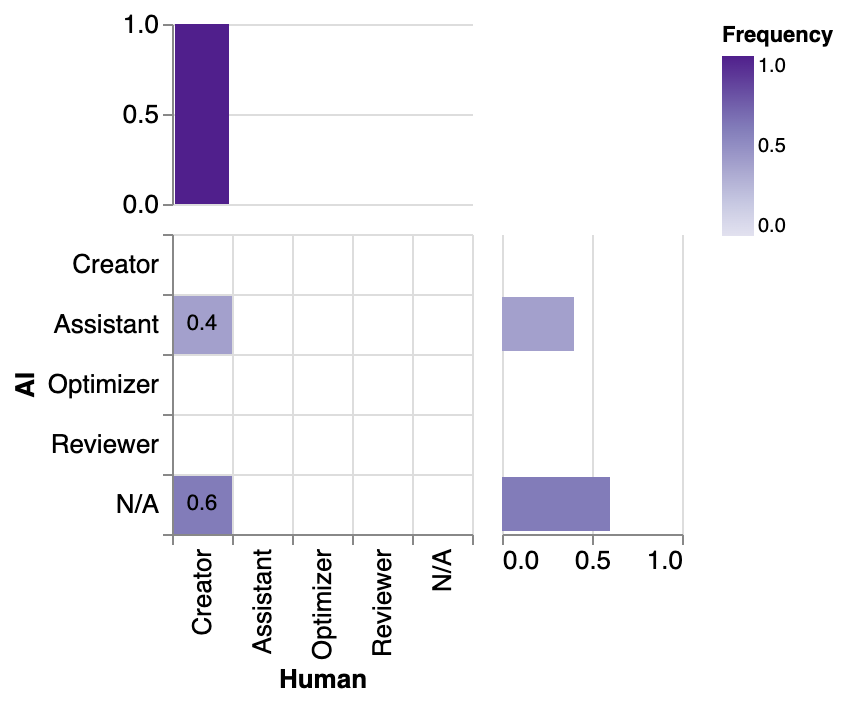}
            \caption{Communication}  
        \end{subfigure}
        \caption
        {The human-AI collaboration pattern frequencies of earlier tools.}
        \label{fig:stage_original}
    \end{figure*}

\begin{figure*}[h!]
        \centering
        \begin{subfigure}[b]{0.24\linewidth}
            \centering
            \includegraphics[width=\textwidth]{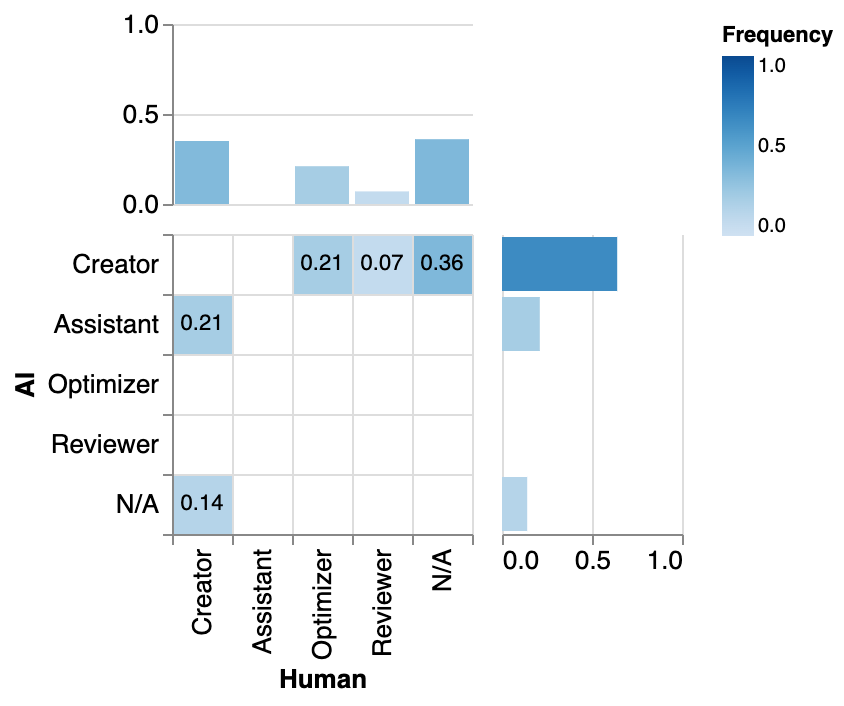}
            \caption{Analysis}  
        \end{subfigure}
        \begin{subfigure}[b]{0.24\linewidth}  
            \centering 
            \includegraphics[width=\textwidth]{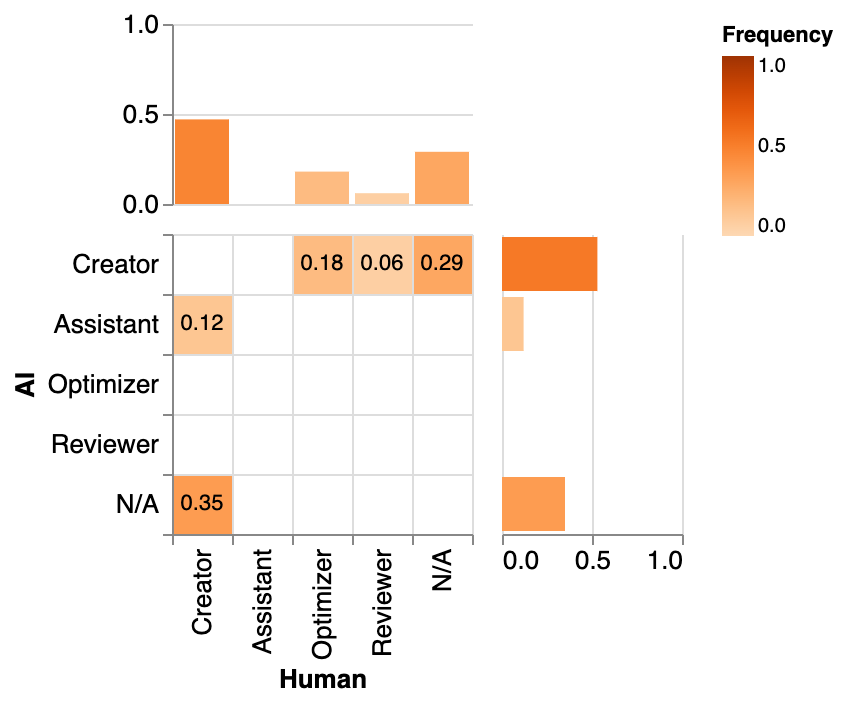}
            \caption{Planning}    
        \end{subfigure}
        \begin{subfigure}[b]{0.24\linewidth}   
            \centering 
            \includegraphics[width=\textwidth]{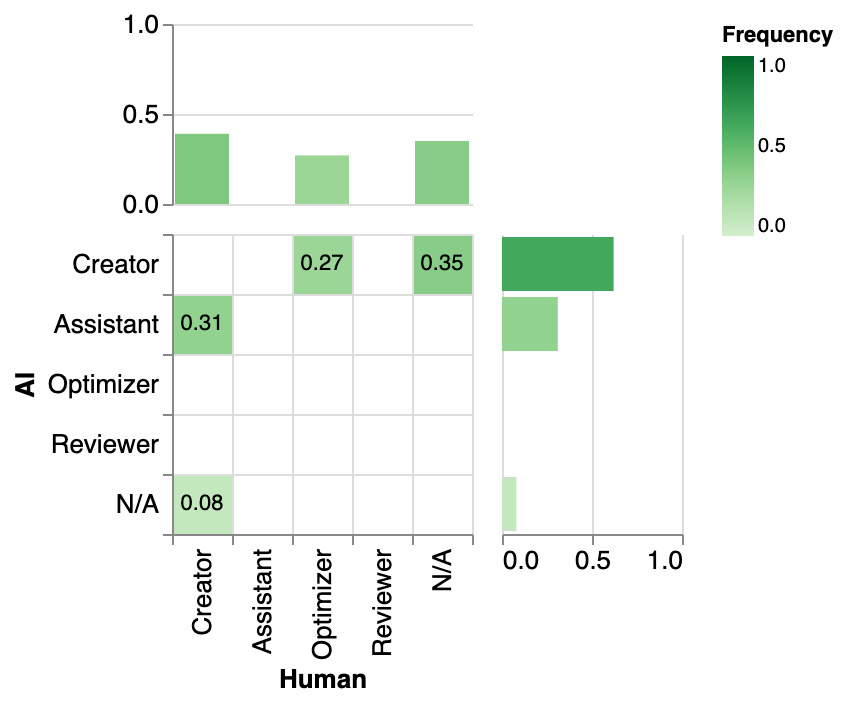}
            \caption{Implementation}    
        \end{subfigure}
        \begin{subfigure}[b]{0.24\linewidth}   
            \centering 
            \includegraphics[width=\textwidth]{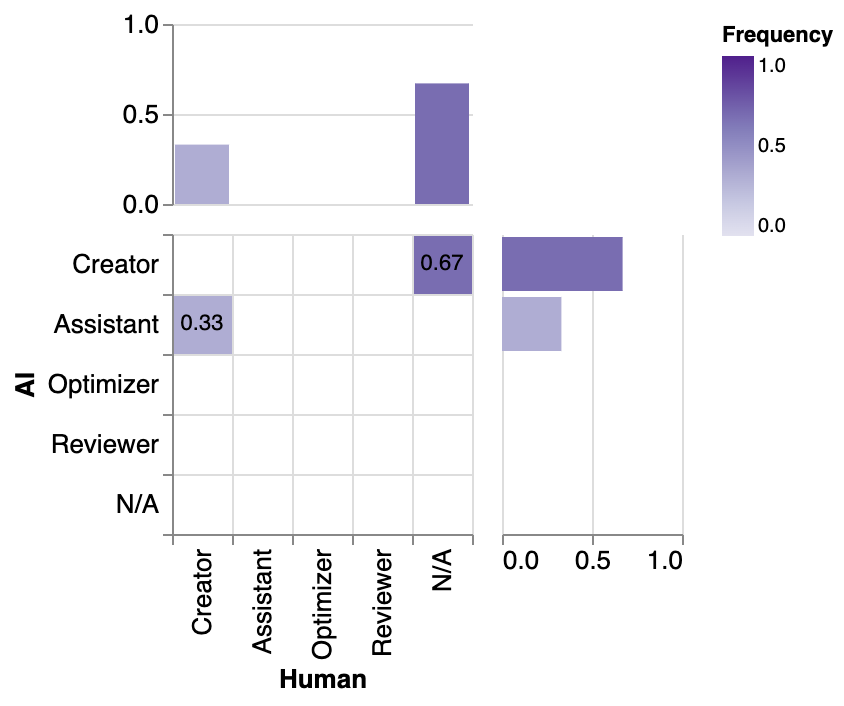}
            \caption{Communication}  
        \end{subfigure}
        \caption
        {The human-AI collaboration pattern frequencies of the latest tools.}
        \label{fig:stage_updated}
    \end{figure*}
